\documentstyle[12pt]{article}
  \def\thebibliography#1{{\bf{References}}\list
 {[\arabic{enumi}]}{\settowidth\labelwidth{[#1]}\leftmargin\labelwidth
   \advance\leftmargin\labelsep
   \usecounter{enumi}}
   \def\newblock{\hskip .11em plus .33em minus -.07em}
   \sloppy
   \sfcode`\.=1000\relax}
  
\begin{document}
\def\ttb{t\bar t}
\def\qqb{Q\bar Q}
\def\gg{\gamma\gamma}
\def\to{\rightarrow}
\def\RM{{R(0)^2\over 2\pi M}}
\def\MeV{{\rm MeV}}
\def\Lms#1{\Lambda_{\overline{MS}}^{(#1)}}
\def\asp{{\alpha_s\over\pi}}
\def\ams{\alpha_{\overline {MS}} }
\def\amsp{{\ams \over\pi}}
\def\eps{\epsilon}
\newcommand{\hoch}[1]{\mbox{\rule[0cm]{0cm}{#1}}}
\everymath={\displaystyle}
\thispagestyle{empty}
\vspace*{-2mm}
\thispagestyle{empty}
\noindent
\hfill TTP92--32\\
\mbox{}
\hfill  Nov.  1992   \\   
\vspace{0.5cm}
\begin{center}
  \begin{Large}
  \begin{bf}
QCD CORRECTIONS TO\\
 TOPONIUM PRODUCTION
 \\AT  HADRON COLLIDERS\footnote{\normalsize Supported by
 BMFT Contract 055KA94P}
   \\
  \end{bf}
  \end{Large}
  \vspace{0.8cm}
  \begin{large}
   J.H. K\"uhn and E.\ Mirkes\\[5mm]
    Institut f\"ur Theoretische Teilchenphysik\\
    Universit\"at Karlsruhe\\
    Kaiserstr. 12,    Postfach 6980\\[2mm]
    7500 Karlsruhe 1, Germany\\
  \end{large}
  \vspace{4.5cm}
  {\bf Abstract}
\end{center}
\begin{quotation}
\noindent
Toponium production at future hadron colliders
is investigated.
Perturbative QCD corrections to the production cross section
for gluon fusion  are calculated
as well as the
contributions from gluon-quark and quark-antiquark collisions to
the total cross section.
The dependence on the  renormalization and factorization scales and
on the choice of the parton distribution functions is explored.
QCD corrections
to the branching ratio of $\eta_{t}$ into $\gamma\gamma$ are
included and the
two-loop QCD potential is used to predict the wave function at
the origin.
The branching
ratio of $\eta_{t}$ into $\gamma Z$, $ZZ$, $HZ$ and $WW$ is
compared
with the $\gamma\gamma$ channel.
\end{quotation}
\newpage
\setcounter{page}{1}
\section{Introduction}
The production of heavy quarks at hadron colliders
has received a lot of
attention from the experimental as well as from the
theoretical side.  Higher
order QCD calculations for open top quark production
\cite{Dawson,Altarelli,Leiden}
provide the theoretical
basis for the current experimental limits \cite{CDF}
on the mass of the top quark.  Given
the current
range for the top mass of 90 to 180 GeV deduced from electroweak
radiative corrections \cite{Carter}
ongoing experimental studies
at the TEVATRON should be able to discover the
top quark in the near future.

Planned experiments at LHC or SSC will lead to
a huge number of $\ttb$ events and
may pin down the top quark mass with a remaining
uncertainty of perhaps 5~GeV \cite{LHC}.
Aside from the intrinsic interest in the precise
top mass determination, the aim
to fix the input in one-loop electroweak
corrections \cite{Schaile} constitutes one of the
important motivations for the drive for
future increased precision in the top
mass.

In \cite{Rubbia} it has been proposed to
decrease the experimental error
significantly through the study of $\eta_t$
in its $\gg$ decay channel.
Making use of the large luminosity of
future hadron colliders, one should be able to
overcome the tiny cross section multiplied
by the small branching ratio into two photons.
The excellent gamma energy resolution, originally
developed for the search for a light Higgs
boson, would in principle allow the
measurement of the mass of the bound state
at about 100 MeV accuracy.  This
would result in a determination of the top
quark mass limited only by
theoretical uncertainties.  It was shown
that the toponium signal could be
extracted from the $\gg$ background
with a signal to noise ratio decreasing
from 2/3 down to 1/8 for $m_t$ between 90 and 120 GeV.

The estimates of \cite{Rubbia} for the production and decay rate
were based on the Born approximation and on
fairly crude assumptions on the bound state  wave
function which was
derived from a Coulomb potential,
neglecting QCD corrections.
The signal
to noise ratio and the statistical significance of the signal as
estimated
in \cite{Rubbia} is just sufficient for the discovery. Therefore a
more precise evaluation of the production rate is mandatory.

In lowest order  the reaction is induced by
gluon-gluon fusion (fig. 1a). QCD
corrections will
be calculated in this work. They affect and indeed increase
the rate in this channel by about 50\%.
 Quark-gluon  scattering and quark-antiquark annihilation into
 $\eta_{t}`$ enter
 at order
 $\alpha_{s}^{3}$. As we shall see the $qg$ process increases the
 production cross
 section by about 8\% whereas the effect of  $q\bar{q}$
annihilation is negligible.

The production cross section for $\eta_{t}$ with
nonvanishing  transverse momentum is
infrared finite
in the order considered here. The total cross section, however,
exhibits
singularities that have to be absorbed in appropriately chosen and
defined parton
distribution functions. The dependence on the {\it factorization
} as
well as {\it renormalization}  scales will be studied as well as the
dependence on
the choice on different parametrizations of parton distribution
functions.
A full NLO
calculation requires the corresponding corrections for the
physically
interesting branching ratio. These are available and will be
implemented.

The bound state
production cross section and the branching ratio of $\eta_{t}$
into $\gamma\gamma$ depend (markedly!)  on $R(0)^{2}$,
the square of the bound state wave
function at
the origin and hence on the potential. The range of $R(0)$
compatible with
the current knowledge of the perturbative two-loop QCD
potential and
the present experimental results for $\Lambda_{QCD}$ will be
explored.

As we shall
see in the following, fairly optimistic assumptions are necessary
to open even
a narrow window for toponium discovery in the mass range of
$m_{t}=90-110$ GeV.
This is the consequence of the dominance single quark
decays (SQD)
and the small branching ratio into two photons. The situation
improves
dramatically if new hypothetical quarks are considered with
suppressed
single quark decay. Examples are $b^{\prime}$ or isosinglet quarks
with small
mixing with ordinary $d,s$ or $b$ quarks. These would dominantly
decay into
gauge and Higgs bosons, leading to spectacular signatures.

The outlay
of this paper is as follows: In section 2 the QCD corrections to
gluon gluon fusion into $\eta_{t}$   and the cross section
for the $q\bar{q}$ and $qG$ initiated processes will be derived.
 In section 3 the formulae for
the decay
rates of $\eta_{t}$ will be listed and the dependence of the wave
function
on the choice of the potential studied. Section 4 contains the
numerical
evaluation of the production cross section and the study
of scale dependencies.
Section 5 contains our conclusions. A brief account of
this work has been presented in \cite{letter}.

\section{QCD corrections to hadronic $\eta_{t}$ production}
For the complete evaluation of NLO corrections
for $\eta_t$ production in gluon
gluon fusion virtual corrections  (fig.~1b)
are required as well as corrections
from real gluon radiation (fig.~1c,d).
In addition
one needs the cross section for quark-antiquark annihilation into
$\eta_{t}+$gluon (fig. 2a) and the (anti-)quark-gluon
(fig.~2b) initiated
reaction into
$\eta_{t}+$quark.
  Technically it is
quite convenient to employ
dimensional regularization for both ultraviolet
and infrared singularities.  The
calculation of amplitudes for bound state
production (or decay) can be performed
in two different ways.  One may either
evaluate directly the {\it amplitude} for the
production of a $(\qqb)$ bound state with the desired
spin and orbital angular momentum
configuration and arrive at an amplitude
proportional to the wave function at
the origin or its derivative \cite{kaplan,guberina}; or alternatively
one may evaluate the {\it production rate}
for open $\qqb$ and subsequently identify
the rate at threshold with the
rate for bound state production, substituting
\begin{equation}
dPS_n(k_{1}+k_{2};k_3,\dots,k_{n},k_Q,k_{\bar Q})
\longrightarrow \RM
dPS_{n-1}(k_{1}+k_{2};k_3,\dots,k_{n},q)
\label{psbound}
\end{equation}
where $M=2m_{t}+E_{Bind}$ and $k_{Q}=k_{\bar{Q}}=q/2$.
For the one particle
final state relevant for resonant $\eta_{t}$ production
in gluon gluon fusion this implies (in $4-2\epsilon$ dimensions):
\begin{equation}
dPS_{2}(k_{1}+k_{2};k_{Q},k_{\bar{Q}})\Rightarrow
\frac{R(0)^{2}}{2\pi M}
2\pi
\delta(s-M^{2})
\end{equation}
and for the $\eta_{t}+$
parton configuration that will be of main concern in
the following
\begin{equation}
dPS_{3}(k_{1}+k_{2};k_{3},k_{Q},k_{\bar{Q}})\Rightarrow
\frac{R(0)^{2}}{2\pi M}
\frac{1}{8\pi s}
\left(\frac{4\pi s}{ut}\right)^{\epsilon}
\frac{1}{\Gamma(1-\epsilon)} \,dt
\label{ps3}
\end{equation}
where the Mandelstam variables are defined by
\begin{equation}
s=(k_{1}+k_{2})^{2}\hspace{1cm}
t=(k_{1}-q)^{2}\hspace{1cm}
u=(k_{2}-q)^{2}=M^{2}-t-s
\end{equation}
In the latter approach it is
important to make sure that
--- in the limit of vanishing relative velocity ---
the cross section
receives solely contributions from the $\qqb$
spin parity configuration corresponding to the bound state.
It is therefore mandatory to
restrict the evaluation of the amplitude to
$\qqb$ color singlet configurations and in the reaction under
consideration to those three gluon states that are totally
antisymmetric with respect to color as well as
to their Lorentz structure.

The former approach leads typically to more compact
expressions\footnote{
See
for instance \cite{Wu} for particularly impressive examples.}.
However, for states
with $J^{PC}=0^{-+}$ the formula for the bound state amplitude
 involves the
$\gamma ^5$ matrix whose formulation in the
context of dimensional regularization
introduces complications.  This purely
formal difficulty is absent in the
second approach, which  for this reason has been adopted in
this calculation.
\subsection{The hadronic cross section}
In this section we define the general structure of the cross sections
for $\eta_{t}$ production in hadronic collisions
\begin{equation}
h_{1}(P_{1})+h_{2}(P_{2})
\rightarrow\eta_{t}+X \rightarrow \gamma+\gamma\,+X
\end{equation}
 within the framework of perturbative QCD.
Here $h_{1},h_{2}$ are unpolarized hadrons with momenta $P_{i}$.
The hadronic cross section in NLO is thus given by
\begin{equation}
\sigma^{H}(S)
=\int dx_{1} dx_{2}
f_{a}^{h_{1}}(x_{1},Q^{2}_{F})f_{b}^{h2}(x_{2},Q^{2}_{F})
 \hat{\sigma}^{ab}(s=x_{1}x_{2}S,\alpha_{s}(\mu^{2}),\mu^{2},Q^{2}_{F})
\end{equation}
where one sums over $a,b=q,\bar{q},g$.
$f_{a}^{h}(x,Q^{2}_{F})$
is the probability density to find parton $a$ with
fraction $x$ in hadron $h$ if it is probed at scale $Q_{F}^{2}$
and $\hat{\sigma}^{ab}$ denotes the parton cross section for the process
\begin{equation}
a(k_{1}=x_{1}P_{1})+b(k_{2}=x_{2}P_{2})\rightarrow \eta_{t}+X
\end{equation}
from which collinear
initial state singularities have been factorized out at a
scale $Q_{F}^{2}$ and implicitly included in the scale-dependent parton
densities $f_{a}^{h}(x,Q_{F}^{2})$.
We next specify the possible partonic subprocesses that contribute to
$\eta_{t}$ production in LO and NLO.
\subsection{Gluon fusion}
In Born approximation the cross section for $gg\rightarrow\eta_{t}$
(fig.~1a) is given by
$(N_{C}=3)$
\begin{equation}
\hat{\sigma}_{Born}^{gg}
=\frac{1}{2s}\frac{1}{64}\frac{1}{4(1-\epsilon)^{2}}
\sum|{\cal{M}}_{Born}|^{2}
\frac{R(0)}{2\pi M}\,2\pi\,\delta(s-M^{2})
\end{equation}
with
\begin{equation}
|{\cal{M}}_{Born}|^{2}=\frac{N_{C}^{2}-1}{4N_{C}}g_{s}^{4}
\,16\,(1-2\epsilon)(1-\epsilon)\,
\end{equation}
The $Q\bar{Q}$
state has been projected onto the color singlet configuration.
The division
by $64\cdot 4\cdot (1-\epsilon)^{2}$ is implied by the average
over $8^{2}$ gluon
colors and $(n-2)^{2}$ gluon helicities in $n$ dimensions.
$g_{s}$ will be replaced by $\alpha_{s}$ through
$g_{s}^{2}=4\pi\alpha_{s}\mu^{2\epsilon}$.
The Born cross section can therefore be cast into the form
\begin{equation}
\hat{\sigma}_{Born}^{gg}
=\frac{\pi^{2}\alpha_{s}^{2}}{3s}\frac{R(0)^{2}}{M^{3}}
\,\mu^{4\epsilon}\,\frac{(1-2\epsilon)}{(1-\epsilon)}\,\,
\delta(1-z)
\end{equation}
where $z=M^{2}/s$.
Virtual
corrections to the rate for $Q\bar{Q}$ production close to threshold
(fig.~1b)
can be combined to
yield\footnote{Here $C_{F}=4/3;\,N_{C}=3;$ and $T_{F}$=1/2.}
\cite{Hagiwara}
\begin{eqnarray}
\sum&2&|{\cal{M}}_{virtual}\ast{\cal{M}}_{Born}|=
\sum|{\cal{M}}_{Born}|^{2}
\,\left(\frac{4\pi\mu^{2}}{M^{2}}\right)^{\epsilon}
\Gamma(1+\epsilon)\nonumber\\
&&
\times\frac{\alpha_{s}}{\pi}\left(
C_{F}\frac{\pi^{2}}{v}+b_{0}\frac{1}{\epsilon_{UV}}
-N_{C}\frac{1}{\epsilon_{IR}^{2}}-b_{0}
\frac{1}{\epsilon_{IR}}+A\right)
\label{virt}
\end{eqnarray}
where
\begin{eqnarray}
A&\equiv&-T_{F}\frac{4}{3}\ln 2+ C_{F}\left(\frac{\pi^{2}}{4}-5\right)
+N_{C}\left(\frac{5}{12}\pi^{2}+1\right)\nonumber\\
b_{0}&\equiv& N_{C}\frac{11}{6}-n_{f}T_{F}\frac{2}{3}
\label{b0def}
\end{eqnarray}
$1/ \epsilon_{UV}$ and $1/ \epsilon_{IR}$ represent poles arising from
ultraviolet and
infrared divergent integrals respectively. The mass singularity
$2\ln(m_{i}/ M)$
in \cite{Hagiwara} has been replaced by $1/ \epsilon_{IR}$.
Since we are considering bound state production the
$1/v$ term ($v$ denotes the relative velocity of the two heavy
quarks) in eq.
(\ref{virt}) is absorbed in the instantaneous potential and
thus effectively dropped.
The QCD coupling
is renormalized in the $\overline{{MS}}$ scheme and is given
by
\begin{equation}
\frac{\alpha_{s}}{\pi}=\frac{\alpha_{\overline{MS}}}{\pi}(\mu^{2})
\left[1-
\frac{\alpha_{\overline{MS}}}{\pi}
b_{0}\left(\frac{1}{\epsilon_{UV}}+\ln 4\pi-\gamma_{E}\right)\right]
\end{equation}
Combining Born
result and virtual corrections one thus obtains for resonant
production
\begin{eqnarray}
\hat{\sigma}_{Born+virtual}^{gg}&=&\sigma_{0}(\epsilon)
\left[1+\frac{\alpha_{s}}{\pi}
\left(\frac{4\pi\mu^{2}}{M^{2}}\right)^{\eps}
\Gamma(1+\epsilon)\,\left(b_{0}\ln\frac{\mu^{2}}{M^{2}}\right.
\right.\nonumber\\
&&\left.\left.
- N_{C}\frac{1}{\epsilon_{IR}^{2}}
- b_{0}\frac{1}{\epsilon_{IR}}
+A\right)\right]\delta\left(1-z \right)
\end{eqnarray}
where the normalization factor
\begin{equation}
\sigma_{0}(\epsilon) =
\frac{1}{s}\frac{\pi^{2}}{3}\frac{R^{2}(0)}{M^{3}}
\alpha^{2}_{\overline{MS}}(\mu^{2})\mu^{4\epsilon}
\frac{(1-2\epsilon)}{(1-\epsilon)}
\end{equation}
has been split off for convenience.

In a second step the cross section for
\begin{equation}
gg\rightarrow gQ\bar{Q} \label{proc1}
\end{equation}
has to be calculated (fig. 1c,d).
Only the kinematical situation will be treated where the
relative momentum
of $Q$ and $\bar{Q}$ vanishes. In an intermediate step the
matrix element for
\begin{equation}
ggg\rightarrow Q\bar{Q} \label{proch}
\end{equation}
will be considered (fig.~1e,f)
and reaction (\ref{proc1})  can be subsequently
obtained  through crossing.
The $Q\bar{Q}$ state will be projected onto a color
singlet configuration through $\delta_{ij}/ \sqrt{3}$. The overall
wave function  of the
gluons is
of course totally symmetric.  Requiring  a totally antisymmetric
configuration
in color space implies, in turn, a totally antisymmetric wave
function in Minkowski space.
This latter
condition secures $J^{PC}=0^{-+}$ for the $Q\bar{Q}$ bound state.
(Conversely,
requiring symmetric wave functions in both color and Minkowski
space would lead to $J^{PC}=1^{--}.)$

The amplitude arising from the diagram fig.~1e is given by:
\begin{eqnarray}
\hat{\cal{A}}(k_{1},k_{2},k_{3};\epsilon_{1},\epsilon_{2},\epsilon_{3})
&=&\nonumber\\
&&\hspace{-3cm} \bar{u}\left((k_{1}+k_{2}+k_{3})/2\right)
        (-ig\not{\epsilon}_{1})
        \frac{i\left[(-\not{k}_{1}+\not{k}_{2}+\not{k}_{3})/2+m\right]
            }{ \left[(-{k}_{1}+{k}_{2}+{k}_{3})^{2}/4-m^{2}\right]}
        (-ig\not{\epsilon}_{2})\nonumber\\
&&\hspace{-3cm}
   \times\frac{i\left[(-\not{k}_{1}-\not{k}_{2}+\not{k}_{3})/2+m\right]
            }{ \left[(-{k}_{1}-{k}_{2}+{k}_{3})^{2}/4-m^{2}\right]}
        (-ig\not{\epsilon}_{3})
        v\left((k_{1}+k_{2}+k_{3})/2\right)
\end{eqnarray}
to be multiplied the color factor
\begin{equation}
\left(\frac{\lambda^{a_{1}}}{2}
      \frac{\lambda^{a_{2}}}{2}
      \frac{\lambda^{a_{3}}}{2}
\right)_{ij}
\frac{\delta_{ij}}{\sqrt{3}}
=
\frac{1}{\sqrt{3}}\frac{1}{4}
\left(i\, f_{a_{1}a_{2}a_{3}}+ d_{a_{1}a_{2}a_{3}}\right)
\end{equation}
As discussed above, only the $f$ term will be retained.
The amplitude
is obtained from the six permutations of the diagram without
triple boson coupling
\begin{equation}
{\cal{A}}=\frac{i}{\sqrt{3}}\frac{1}{4}\,f_{a_{1}a_{2}a_{3}}
\sum_{i,j,m} \epsilon_{ijm}
\hat{\cal{A}}(k_{i},k_{j},k_{m};
\epsilon_{i},\epsilon_{j},\epsilon_{m})
\end{equation}
The second type of  amplitues, represented by fig.~1f,
involve the triple gluon vertex. Their sum reads as follows
\begin{eqnarray}
\hat{\cal{B}}(k_{1},k_{2},k_{3};\epsilon_{1},\epsilon_{2},\epsilon_{3})
&=&\nonumber \\
&&\hspace{-4cm}\bar{u}\left((k_{1}+k_{2}+k_{3})/2\right) \left[
        (-ig\gamma_{\alpha})
        \frac{i\left[(-\not{k}_{1}-\not{k}_{2}+\not{k}_{3})/2+m\right]
            }{ \left[(-{k}_{1}-{k}_{2}+{k}_{3})^{2}/4-m^{2}\right]}
        (-ig\not{\epsilon}_{3})\nonumber\right.\\
&&\hspace{-4cm} \left.  +\,
         (-ig\not{\epsilon}_{3})
      \frac{i\left[(\not{k}_{1}+\not{k}_{2}-\not{k}_{3})/2+m\right]
            }{ \left[({k}_{1}+{k}_{2}-{k}_{3})^{2}/4-m^{2}\right]}
        (-ig\gamma_{\alpha})  \right]
        v\left((k_{1}+k_{2}+k_{3})/2\right)\label{ggg}\\
&&\hspace{-4cm} \times \frac{-i}{(k_{1}+k_{2})^{2}} (-g)
     \left\{\hoch{4mm} (k_{1}-k_{2})^{\alpha}\epsilon_{1}\epsilon_{2}
         +   (k_{1}+2k_{2})\epsilon_{1}\epsilon_{2}^{\alpha}
         +   (-2k_{1}-k_{2})\epsilon_{1}^{\alpha}\epsilon_{2}\right\}
\nonumber
\end{eqnarray}
with a color factor
\begin{equation}
\left(\frac{\lambda^{b}}{2}
      \frac{\lambda^{a_{3}}}{2}
\right)_{ij}
\frac{\delta_{ij}}{\sqrt{3}}
f_{a_{1}a_{2}b}
=
\frac{1}{\sqrt{3}}\frac{1}{2}
 f_{a_{1}a_{2}a_{3}}
\end{equation}
Three cyclic permutations  are combined in this class of
diagrams
\begin{equation}
{\cal{B}}=\frac{i}{\sqrt{3}}\frac{1}{4}\,f_{a_{1}a_{2}a_{3}}
\sum_{i,j,m (cyclic)}
\frac{2}{i}
\hat{\cal{B}}(k_{i},k_{j},k_{m};\epsilon_{i},\epsilon_{j},\epsilon_{m})
\end{equation}

The ghost amplitude ${\cal{C}}$ is easily obtained from eq.~(\ref{ggg})
through the replacement of the curly bracket $\{\ldots\}\rightarrow
k_{2}^{\beta}$, where $k_{1},k_{2}$ and $k_{3}$ denote the momenta of
the
incoming ghost, antighost and gluon respectively.

The squared
matrix element is finally obtained in $n=4-2\epsilon$ dimensions
\begin{eqnarray}
\sum|{\cal{M}}|_{R}^{2}&=& \sum|{\cal{A}}+{\cal{B}}|^{2}
-2\sum|{\cal{C}}|^{2}=
\sum_{a_{1},a_{2},a_{3}} \frac{1}{48}\,f_{a_{1}a_{2}a_{3}}^{2}\, g^{6}\,
 128(1-\epsilon)
\label{matalleps}
\\
&&\hspace{-3cm}\times
\left(\frac{st+tu+us}{(s-M^{2})(t-M^{2})(u-M^{2})}\right)^{2}
\left[\frac{M^{8}+s^{4}
+t^{4}+u^{4}}{stu}(1-2\eps)+4\epsilon M^{2}\right]
\nonumber
\end{eqnarray}
where
\begin{equation}
s=(k_{1}+k_{2})^{2}\hspace{1cm}
t=(k_{1}+k_{3})^{2}\hspace{1cm}
u=(k_{1}+k_{3})^{2};\hspace{1cm}
M=2m+E_{Bind}
\end{equation}
After crossing
($k_{3}\rightarrow -k_{3}$) this result is applicable to the
reaction of interest (\ref{proc1}).
In the $Q\bar{Q}$
threshold region (with the color restriction discussed above)
one obtains
\begin{equation}
d\sigma(gg\rightarrow gQ\bar{Q})=
\frac{1}{2s}
\underbrace{
\frac{1}{64}
\frac{1}{4(1-\epsilon)^{2}}\sum|{\cal{M}}|^{2}_{R} }_{:=g^{6}
\overline{|{\cal{M}}|^{2}_{R}}}
\,dPS_{3}
\end{equation}
where gluon helicities and colors were averaged as usual.
Replacing the three-  by the two-particle phase space as indicated in
eqs.~(\ref{psbound}) and (\ref{ps3})  one obtains
\begin{eqnarray}
d\sigma(gg\rightarrow g\eta_{t})&=&
\sigma_{0}(\eps)\,\frac{\alpha_s}{\pi}\frac{3}{2}\frac{M^{2}}{s}
\label{matall} \\
&&\hspace{-3cm}\times
\left(\frac{st+tu+us}{(s-M^{2})(t-M^{2})(u-M^{2})}\right)^{2}
\frac{M^{8}+s^{4}+t^{4}+u^{4}}{stu}
\left(\frac{4\pi s\mu^{2}}{ut}\right)^{\epsilon}
\frac{dt}{\Gamma(1-\epsilon)}\nonumber
\end{eqnarray}
The term proportional $4\epsilon M^{2}$ in eq.~(\ref{matalleps})
without singular denominator  will not contribute in the limit
$\epsilon\rightarrow 0$ and has therefore
been dropped.
In the limit
$\epsilon\rightarrow 0$ the result coincides with eq. (8.45) of
\cite{Wu}.
For the subsequent
discussion it will be convenient to follow closely the
treatment of NLO corrections for the
Drell-Yan process presented in the
classical paper by G.Altarelli et al. \cite{aem}.
The invariants
$s,t$ and $u$ are expressed in terms of $M^{2}, y$ and $z$
by
\begin{equation}
s=\frac{M^{2}}{z};\hspace{1cm}
t=-\frac{M^{2}}{z}(1-z)(1-y);\hspace{1cm}
u=-\frac{M^{2}}{z}(1-z)y
\end{equation}
and one arrives at
Gluon\begin{eqnarray}
\sigma(gg\rightarrow g(Q\bar{Q}))
&=&\sigma_{0}(\eps)\,\frac{\alpha_{s}}{\pi}\Gamma(1+\eps)
\left(\frac{4\pi\mu^{2}}{M^{2}}\right)^{\epsilon}\nonumber \\
&&\hspace{-3cm}\times \frac{6}{\Gamma(1+\epsilon)\Gamma(1-\epsilon)}
\int_{0}^{1}f^{gg}z^{\epsilon}(1-z)^{1-2\epsilon}y^{-\epsilon}
(1-y)^{-\epsilon} \,dy
\label{sggg}
\end{eqnarray}
The normalization
factors have been chosen to allow for easy combination
with the virtual corrections and $f^{gg}$ is given by
\begin{eqnarray}
f^{gg}&=&
\frac{(1-\eps)}{(1-2\eps)}\,M^{2}\overline{|{\cal{M}}|^{2}_{R}}
\nonumber\\
&=&
\frac{z}{4}
\left(\frac{(1-z)(1-y)y-1}{(1-z)[1-y(1-z)][(1-z)y+z]}\right)^{2}
\nonumber \\
&\times&\left[z^{4}+1 +(1-z)^{4}((1-y)^{4}+y^{4})\right]
\left(\frac{1}{y}+\frac{1}{1-y}\right)
\end{eqnarray}
As a consequence
of the symmetry of $f^{gg}$ with respect to $y\leftrightarrow
(1-y)$
it suffices
to evaluate the $1/y$ term. The $y$ integration can be performed
with the help of the distribution identities
\begin{eqnarray}
x^{-1-\eps}&=&-\frac{1}{\eps}\delta(x)+\left(\frac{1}{x}\right)_{+}
-\eps\left(\frac{\ln x}{x}\right)_{+}\nonumber\\
z^{\eps}(1-z)^{-1-2\eps}&=&
-\frac{1}{2\eps}\delta(1-z)+\left(\frac{1}{1-z}\right)_{+}
+\eps\frac{\ln z}{1-z}-2\eps
\left(\frac{\ln(1-z)}{(1-z)}\right)_{+}\nonumber\\
y^{-1-\eps}(1-y)^{-\eps}&=&
-\frac{1}{\eps}\delta(y)+\left(\frac{1}{y}\right)_{+}
-\eps\,\,\frac{\ln(1-y)}{y}-\eps\left(\frac{\ln y}{y}\right)_{+}
\end{eqnarray}
and
\begin{equation}
\frac{1}{\Gamma(1-\eps)}
\frac{1}{\Gamma(1+\eps)}=1-\eps^{2}\frac{\pi^{2}}{6}+\ldots
\end{equation}
Combining the Born
result with real radiation and virtual corrections one
obtains
\begin{eqnarray}
\sigma(gg\rightarrow\eta_{t}+X)&=&\sigma_{0}(\eps)\left\{\delta(1-z)
+\frac{\alpha_{s}}{\pi}\left(\frac{4\pi\mu^{2}}{M^{2}}\right)^{\eps}
\Gamma(1+\eps)   \right.\nonumber\\
&&\hspace{-3cm}\times\left.\left[
\left(b_{0}\ln\frac{\mu^{2}}{M^{2}}-N_{C}
\frac{\pi^{2}}{3}+A\right)\delta(1-z)
-\frac{1}{\eps}P_{gg}(z)+N_{C}F(z)\right]\right\}
\end{eqnarray}
where $b_{0}$ and $A$ are given in eqs.~(\ref{b0def})  and
\begin{eqnarray}
F(z)&=&
     \Theta(1-z)
    \left[\,\,\,\,
    \frac{11z^{5}+11z^{4}+13z^{3}+19z^{2}+6z-12}{6z(1+z)^{2}}\right.
      \nonumber \\[2mm]
    &&\hspace{-3cm}
    \,+\,\,\,\,4\left(\frac{1}{z}+z(1-z)-2\right)\ln(1-z)
     +\,\,\,4\,\left(\frac{\ln(1-z)}{1-z}\right)_{+}
 \label{sig} \\[2mm]
    &&\hspace{-3cm}
      \left.
     +\,\,\left(\frac{2(z^{3}-2z^{2}-3z-2)
     (z^{3}-z+2)z\,\ln(z)}{(1+z)^{3}(1-z)}
             -3\right)\frac{1}{1-z}\,\right]\nonumber
\end{eqnarray}
and
\begin{equation}
P_{gg}=2\,N_{C}\left(\frac{1}{z}
+\left(\frac{1}{1-z}\right)_{+}+z(1-z)-2\right)
+b_{0}\delta(1-z);
\end{equation}
The $1/ \eps$ term
can be absorbed in the gluon densities. If these are defined
at a factorization
scale $Q^{2}_{F}$, one effectively subtracts from the above
result the term
\begin{equation}
\sigma_{0}(\eps)\frac{\alpha_{s}}{2\pi}
\left(\frac{4\pi\mu^{2}}{Q^{2}_{F}}\right)^{\eps}\Gamma(1+\eps)
     \left(-\frac{1}{\eps}P_{gg}(z)+C_{gg}(z)\right)
\end{equation}
In the $\overline{MS}$ scheme $C_{gg}=0$, in the DIS scheme
\begin{equation}
C_{gg}(z)=-n_{f}\left[\left(z^{2}+(1-z)^{2}\right)
\ln\left(\frac{1-z}{z}\right)
            +8z(1-z)-1 \right]
\end{equation}
The partonic cross section is then given by
\begin{eqnarray}
\hat{\sigma}^{gg}&=&
\frac{1}{s}\frac{\pi^{2}}{3}\frac{R(0)^{2}}{M^{3}}
\alpha_{\overline{MS}}^{2}(\mu^{2})
\left\{\hoch{6mm}\delta(1-z)\right.\nonumber\\
&&\hspace{-1.5cm}
     +\, \frac{\alpha_{\overline{MS}}(\mu^{2})}{\pi}
\, \left\langle \,\,\delta(1-z)\,\left(
       b_{0}\ln\frac{\mu^{2}}{M^{2}}+
      N_{C}\left(1+\frac{\pi^{2}}{12}\right)
                      +C_{F}\left(\frac{\pi^{2}}{4}-5\right)
                      -\frac{4}{3}T_{f}\ln(2) \right) \right.
                                \nonumber     \\[2mm]
     &&\hspace{-0.5cm}
\left.  -\,\,\ln\frac{Q^{2}_{F}}{M^{2}}P_{gg}(z)-C_{gg}(z)+N_{C}\,
F(z) \right\rangle
                 \Biggr\}.
\end{eqnarray}
\subsection{Gluon quark scattering}
Gluon quark scattering
$g(k_{1})+q(k_{2})\rightarrow q(k_{3})+\eta_{t}(P)$
proceeds evidently
in leading order $\alpha_{s}^{3}$ through the diagrams in
fig.~2b.
Collinear singularities
have to be absorbed in the structure functions. The
(dimensionally regularized)
squared matrix element for the production of
$Q\bar{Q}$ at threshold is given by
\begin{equation}
|{\cal{M}}^{gq}|^{2}(s,t,u)=32\left[
-\frac{s^{2}+u^{2}}{(s+u)^{2}}
\frac{1}{t}(1-2\eps)+\frac{\eps}{t}\right]
\end{equation}
and has to be supplemented by the color factor
\begin{equation}
\sum\left|\frac{1}{\sqrt{3}}\delta_{ij}
\left(\frac{\lambda^{a}}{2}
\frac{\lambda^{b}}{2}\right)_{ij}
\left(\frac{\lambda^{b}}{2}\right)_{kl}\right|^{2}
=\frac{1}{3}
\end{equation}
and the coupling constant $g_{s}^{6}$.
After spin and color averaging
the cross section can then be cast into a form
identical to eq.~({\ref{sggg}}) with $f^{gg}$ replaced by
\begin{equation}
f^{gq}=\frac{1}{9(1-z)}\frac{z}{(1-y)}\left[
\frac{1+y^{2}(1-z)^{2}}{(1-y(1-z))^{2}}-\frac{\eps}{1-2\eps}\right]
\end{equation}
The $y$ integration
is performed as before and only a single pole remains
\begin{eqnarray}
\sigma^{gq}&=&\sigma_{0}(\eps)
\frac{\alpha_{s}}{2\pi}\Gamma(1+\eps)
\left(\frac{4\pi\mu^{2}}{M^{2}}\right)^{\eps}
\left\{-\frac{1}{\eps}P_{gq}(z)\right. \nonumber \\
&&\left.
+\,\,\ln\left(\frac{(1-z)^{2}}{z}\right)P_{gq}(z)+
C_{F}\frac{2(1-z)}{z}(1-\ln z)
+C_{F}\,z
\right\}
\end{eqnarray}
with
\begin{equation}
P_{gq}(z)=C_{F}\frac{1+(1-z)^{2}}{z}
\end{equation}
The $1/ \eps$ term
can again be absorbed by the (quark) structure function.
This amounts effecively to the subtraction of
\begin{equation}
\sigma_{0}(\eps)\frac{\alpha_{s}}{2\pi}
\left(\frac{4\pi\mu^{2}}{Q_{F}^{2}}\right)^{\eps}
\Gamma(1+\eps)\left(-\frac{1}{\eps}P_{gq}(z)+C_{gq}(z)\right)
\end{equation}
from the above result.
The function $C_{gq}$ is again specific to the
subtraction scheme:
$C_{gq}$ vanishes
is the $\overline{MS}$ scheme and in the DIS scheme
it is given by
\begin{equation}
C_{gq}(z)=-C_{qq}(z)
\end{equation}
where
\begin{equation}
  \begin{array}{lll}
C_{qq}(z)&=&C_{F}\left[
(1+z^{2})\left(\frac{\ln(1-z)}{1-z}\right)_{+}
-\frac{3}{2}\frac{1}{(1-z)_{+}}
-\frac{1+z^{2}}{1-z}\ln z    \right.
\\[5mm]
&& \left.
\hspace{2cm}+3+2z-\left(\frac{9}{2}+\frac{1}{3}\pi^{2}\right)
\,\delta(1-z)\right]
   \end{array}
  \end{equation}
One therefore  arrives
at the following formula for the parton cross section
  \begin{eqnarray}
\hat{\sigma}^{gq}(z)&=&
\sigma_{0}(0)\frac{\alpha_{s}}{\pi}\Theta(1-z)\left[
\frac{1}{2}P_{gq}(z)
\left(\ln \frac{M^{2}(1-z)^{2}}{Q^{2}_{F}\,z}+1\right)
\right.\nonumber\\
&&\left.-C_{F}\frac{1-z}{2z}\ln z-\frac{1}{2}C_{gq}(z)\right]
  \end{eqnarray}
The $q(k_{1})+g(k_{2})$
and $g(k_{1})+q(k_{2})$ initiated reactions can
evidently be deduced
from the same partonic cross section, as well as the
corresponding
antiquark processes. All these will be included in our numerical
analysis.
\subsection{Quark-antiquark annihilation}
Quark-antiquark annihilation (fig.~2a) can be obtained
 in a straightforward manner by
crossing from the previous calculation:
  \begin{equation}
|{\cal{M}}^{q\bar{q}}|^{2}(s,t,u)
=|{\cal{M}}^{gq}|^{2}(u,t,s)\Rightarrow 32
\frac{u^{2}+t^{2}}{(u+t)^{2}s}
  \end{equation}
No infrared divergency
arises and hence $\eps$ can be put at zero from the
beginning.
The angular integration is trivial and one finds
  \begin{equation}
\hat{\sigma}^{q\bar{q}}(z)=\sigma_{0}(0)
\frac{\alpha_{s}}{\pi}\frac{32}{27}z(1-z)
  \end{equation}
This  completes  the evaluation of parton cross sections.
\section{Bound state properties}
The decay of $\eta_{t}$
are completely dominated by the single quark decay mode
\cite{acta,Bigi}. Decays
into two photons constitute a tiny fraction of all
events and the
branching ratio is to a very good approximation given by
  \begin{equation}
BR(\gamma\gamma)=\frac{\Gamma_{\gamma\gamma}}{\Gamma_{SQD}}
  \end{equation}
Up to small boundstate
 corrections $\Gamma_{SQD}$ is given by twice the decay rate
of a top quark $\Gamma_{t}$.
For a complete
NLO calculation of $\eta_{t}$ production and decay into
$\gamma\gamma$
the $QCD$ corrections to the decay rates must be included. In
Born approximation
\begin{eqnarray}
\Gamma_{t}\left(\mbox{Born} \right) &=&\nonumber \\
&&\hspace{-2cm}
\frac{G_F m^3_t}{8\sqrt 2\pi}\,
\frac{2p_W}{m_t}
  \left\{ \left[1-\left(\frac{m_b}{m_t}\right)^2\right]^2
+ \left[ 1 + \left( \frac{m_b}{m_t} \right)^2 \right]
\left( \frac{m_W}{m_t}\right)^2
- 2 \left( \frac{m_W}{m_t} \right)^4 \right\}
\label{eq:1}
\end{eqnarray}
QCD corrections lead to a reduction by about 7\%:
$$\Gamma_t=\Gamma_t(\mbox{Born})(1-{2\over3}\asp f)$$
The complete formula for $f$ is
given in \cite{Jezk}.  For the decay rate into a real $W$,
neglecting its width as
well as the mass of the bottom quark,
one obtains \cite{Jezk}
\begin{eqnarray}
f & = & {\cal F}_1 / {\cal F}_0 \nonumber \\
{\cal F}_0 & = & 2(1-y)^2 (1+2y) \nonumber \\
{\cal F}_1 & = &
{\cal F}_0 \left[ \pi ^2 + 2 Li_2 (y) - 2Li_2 (1-y)\right]
\nonumber \\
 & & + 4y(1-y-2y^2) \ln y + 2(1-y)^2 (5+4y) \ln(1-y) \nonumber \\
 & & - (1-y)(5+9y-6y^2)
\label{eq:21}
\end{eqnarray}
where $y=m^2_W / m^2_t$, which is a good
approximation to the full answer.
The branching
ratio into $\gg$ is  therefore given to high accuracy by
\begin{equation}
Br(0^{-+}\to\gg)={1\over\Gamma_{SQD}}
 12 Q_t^4\alpha^2 \frac{|R(0)|^{2}}{M^2}
 \Biggl[1 + \asp\Biggl(\frac{\pi^2}{
 3} - \frac{20}{3}\Biggr)\Biggr]
\end{equation}
The QCD corrections
are negative in  numerator and denominator and cancel to a
large extent in the branching ratio.
This completes our
list of formulae for $\eta_t$ production and decay.

As stated in the introduction, the most important
ingredients for the numerical
evaluation are the wave function at the origin
and the gluon distribution.  The
former are calculated\footnote{The numerical evaluation of $R(0)$
has been performed with the program BOUNDL \cite{Boundl}.}
for the two-loop potential $V_J$ \cite{Igi}.
The results for the dimensionless quantity $R(0)^2/M^3$
are displayed in fig.~3a
 for the
potentials with the  $\Lms{4}$ values of 200, 300 and
500 MeV, corresponding
to $\ams(M_Z) =0.109, 0.1165$ and 0.127 respectively
 (and corresponding to
 $\Lms{5}= 0.132, 0.207$ and 0.366 MeV).
This choice for $\Lms{4}$ covers well the present range
$\ams(M_{Z})=0.117\pm 0.07$
derived from a global fit to the present data
\cite{Altarelli}.
In the subsequent
discussions the results for $V_{J}$ with the  values
of 200 and 500 MeV for
$\Lambda$ will be considered as indicative for the uncertainty from
the wave
function.
The branching ratio of $\eta_{t}$ into $\gamma\gamma$ is shown in
fig.~3b.

In principle decays
into massive gauge and Higgs bosons could be of interest
and have been
calculated in \cite{acta,barger,kurep}\footnote{The result for
$\Gamma_{ZZ}$
in \cite{kurep} (eq. 9.1c) should be multiplied by a factor
$(v_{a}^{2}+a_{a}^{2})/16.$}.
The decay rate is dominated entirely by the single quark decay.
The branching
ratio of $\eta_{c}$ into $\gamma Z$ (and correspondingly for
$HZ,ZZ$ and $WW$
is therefore given by the branching ratio into $\gamma\gamma$
as calculated before, multiplied by the ratio $\Gamma_{\gamma Z}/
\Gamma_{\gamma\gamma }$.
As shown in fig.~3c these latter ratios vary between
0.2 (for $\gamma Z$) and 40 (for $HZ$) in the mass range of interest.

\section{Numerical results}
With these ingredients
it is straightforward to arrive at numerical predictions
for the production
cross section. If not stated otherwise, these will be based
on the parton
distribution functions of MT (Morfin-Tung) set B1 \cite{MT}.
$\alpha_{s}$
is chosen accordingly to the $\Lambda$ value consistent with the
parton distribution functions with five flavours.
All curves correspond to the DIS factorization scheme.

To investigate the stability of the predictions in a first step
 the dependence
 of the results on the renormalization scale $\mu$ and the
 factorization scale $Q_{F}$ will be explored.
The resonance mass will be kept fixed at  240
GeV  and $\mu$ and $Q_{F}$ will be chosen identically.
The result
of the variation for the $gg$ induced process alone is compared
to the LO predictions in fig.~4a, the sum of all contributions in
fig.~4b for $\sqrt{S}$=16  and 40 TeV.
Varying $\mu=Q_{F}$
in the range $M/2$  to $2M$ induces an uncertainty of about
$\pm 8\%$ for SSC and  LHC.
These curves have been obtained with $V_{J}(200 \mbox{MeV})$.
In the
following discussion, both $\mu$ and $Q_{F}$ will be fixed to $M$.

The magnitude
of the QCD corrected cross section including all subprocesses
is compared to the Born cross section in fig.~5(a) for $\sqrt{S}$=
16 and 40 TeV and (b) for $\sqrt{S}=1.8 $ TeV
The size of
the corrections amounts to 40-50\% for LHC and SSC energies,
whereas the corrections for Tevatron energies amounts to 80-90\%.

 To give an idea about the importance of different partonic
 processes,
 their relative contribution to the production rate is displayed in
 fig.~6.
As anticipated
in \cite{letter} $\eta_{t}$ production is dominated by gluon
fusion with the
remaining contributions amounting about 10\%  at $\mu=Q_{F}=M$.
This holds true for all quark masses and energies of interest.

To study the dependence of the cross section on the choice of parton
distribution functions,
the predictions based on a variety of parametrizations
(MT set SN \cite{MT}, DFLM \cite{DFLM}, MRS set B200 \cite{MRS},
KMRS set B \cite{KMRS}, GRV \cite{GRV})
are compared   in fig.~7 to those based on  MT set B1.
$\alpha_{s}$ is
chosen accordingly to the $\Lambda$ value consistent with the
parton distribution functions with five flavours.
The resulting difference amounts to less than 25\% and is in any case
small compared
to the ambiguity from different choices of $\Lambda$ in the
potential.

The final predictions for the cross section are shown in fig.~8a
for LHC and SSC and in fig.~9 for 1.8 TeV.
The solid line corresponds to $V_{J}(\Lms{4}=300 \mbox{MeV})$, parton
distribution
functions of MT set B1 and $\alpha_{s}$ derived from $\Lms{5}$.
The dashed and dotted
curves enclose the range of predictions resulting from
the
dominant uncertainty
in the potential and correspond to $V_{J}(\Lms{4}=500
\mbox{MeV})$  to $V_{J}(\Lms{4}=200 \mbox{MeV})$.
The corresponding predictions for the cross section
times the branching ratio into $\gamma\gamma$ are displayed
in fig.~8b.

The results are significantly below those of \cite{Rubbia}
even for an optimistic choice of the strong coupling constant.\\[1cm]
{\bf Acknowledgements:}\\[5mm]
We would like to
thank P.M. Zerwas for stimulating discussions which inititated
this work
and for informations about details of \cite{Zerwas}.
We thank  M.~Jezabek for providing the program BOUNDL.\\[4cm]
\def\app#1#2#3{{\it Act. Phys. Pol. }{\bf B #1} (#2) #3}
\def\apa#1#2#3{{\it Act. Phys. Austr.}{\bf#1} (#2) #3}
\def\lhc{Proc. LHC Workshop, CERN 90-10}
\def\npb#1#2#3{{\it Nucl. Phys. }{\bf B #1} (#2) #3}
\def\plb#1#2#3{{\it Phys. Lett. }{\bf B #1} (#2) #3}
\def\prd#1#2#3{{\it Phys. Rev. }{\bf D #1} (#2) #3}
\def\prl#1#2#3{{\it Phys. Rev. Lett. }{\bf #1} (#2) #3}
\def\prc#1#2#3{{\it Phys. Reports }{\bf C #1} (#2) #3}
\def\cpc#1#2#3{{\it Comp. Phys. Commun. }{\bf #1} (#2) #3}
\def\nim#1#2#3{{\it Nucl. Inst. Meth. }{\bf #1} (#2) #3}
\def\pr#1#2#3{{\it Phys. Reports }{\bf #1} (#2) #3}
\def\sovnp#1#2#3{{\it Sov. J. Nucl. Phys. }{\bf #1} (#2) #3}
\def\jl#1#2#3{{\it JETP Lett. }{\bf #1} (#2) #3}
\def\jet#1#2#3{{\it JETP Lett. }{\bf #1} (#2) #3}
\def\zpc#1#2#3{{\it Z. Phys. }{\bf C #1} (#2) #3}
\def\ptp#1#2#3{{\it Prog.~Theor.~Phys.~}{\bf #1} (#2) #3}
\def\nca#1#2#3{{\it Nouvo~Cim.~}{\bf #1A} (#2) #3}
\newpage
\sloppy
\raggedright

\vspace{2cm}
\noindent
{\bf Figure captions}\\[2mm]
\begin{itemize}
\item[{\bf Fig. 1}]
(a) Lowest order diagrams for $eta_{t}$ production,\\
(b) Diagrams contributing from virtual corrections, \\
(c,d) Diagrams contributing to corrections from gluon emission,\\
(e,f) Diagrams for $ggg\rightarrow Q\bar{Q}$.
\item[{\bf Fig. 2}]
Diagrams contributing
to corrections from (a) $q\bar{q}\rightarrow\eta_{t} q$
and\\
 (b)  $qg\rightarrow\eta_{t} g$
gluon emission.
\item[{\bf Fig. 3.}]
 (a) Predictions for
$R(0)^2/M^3$ of the S--wave ground state and \\
(b) branching ratio of
$\eta_{t}\rightarrow \gamma\gamma$
as  functions of $M$ for the potential $V_{J}$ with
$\Lms{4}$=200 (solid), 300 (dotted) and 500 MeV (dashed).\\
(c)
Decay rates of $\eta_{t}$
into $\gamma Z$, $ZZ$  $WW$ and $HZ$ (for $m_{H}=70$
GeV) normalized
to the $\gamma\gamma$ channel.
\item[{\bf Fig 4.}]
(a) Scale dependence
 for $pp\rightarrow
 \eta_{t}+X$  for the $gg$ induced process alone and
(b) scale dependence
for all contributions. Both LO (dashed) and NLO
(solid) results are shown for $\sqrt{S}$=16 and 40 TeV.
 We use MT set B1 parton distributions with
$\Lms{4}$=194 MeV
for four flavours and work in the DIS factorization
scheme.
\item[{\bf Fig 5.}]
(a) Ratio between the radiatively corrected cross section
 for $pp\rightarrow \eta_{t}+X$
and the lowest order result
for $\sqrt{S}$=16 and 40 TeV and (b)
 ratio between the radiatively corrected cross section
 for $p\bar{p}\rightarrow \eta_{t}+X$ and
 the lowest order result
for $\sqrt{S}$=1.8 TeV.
 We use MT set B1 parton distributions with
$\Lms{4}$=194 MeV for four flavours and work in the DIS factorization
scheme.
\item[{\bf Fig 6.}]
Relative contributions to $\eta_{t}$ production at $\mu=Q_{F}=M$
for $\sqrt{S}$=16 (dotted) and 40 (solid) TeV.
LO contributions are (A) $gg\rightarrow \eta_{t}+X$,
NLO contributions are from (B) $gg\rightarrow \eta_{t}+g+X$,
(C) $gq\rightarrow\eta_{t}+q+X$,
(D) ($q\bar{q}\rightarrow\eta_{t}+g+X )\times 100$.
\item[{\bf Fig 7.}]
Prediction for the production cross section of
$\eta_{t}$ at LHC
using different sets of parton distributions functions
(MT set SN \cite{MT}, DFLM \cite{DFLM}, MRS set B200 \cite{MRS},
KMRS set B \cite{KMRS} , GRV\cite{GRV})
normalized to the predictions obtained with MT set B1 \cite{MT}.
\item[{\bf Fig 8.}]
(a) Cross section for $\eta_{t}$ production
including NLO corrections
and (b) cross section
multiplied by the two photon branching ratio
at $\sqrt{S}=16$ and 40 TeV for the
 potential $V_{J}$ with
$\Lms{4}$=300 (solid),
$\Lms{4}$=200 (dotted)
 and $\Lms{4}$=500 MeV (dashed).
\item[{\bf Fig 9.}]
Same as fig. 8a for 1.8 TeV.
\end{itemize}
\end{document}